\documentstyle[psfig]{article}
\newcommand{\gtequiv}{\lower2pt\hbox{$\:\stackrel{>}{
\scriptstyle\sim}\:$}}
\newcommand{\ltequiv}{\lower2pt\hbox{$\:\stackrel{<}{
\scriptstyle\sim}\:$}}

\begin{document}
\noindent
{\small Published in:
{\em Single-Electron Tunneling and Mesoscopic Devices}, edited by H. Koch and
H. L\"{u}bbig (Springer, Berlin, 1992): pp.\ 175--179.}\bigskip\\

\noindent {\Large\bf Resonant Josephson Current\\ through a Quantum
Dot}\bigskip\\ C. W. J. Beenakker$^{a,b}$ and H. van
Houten$^{a}$\bigskip\\ $^{a}$Philips Research Laboratories, 5600 JA
Eindhoven, The Netherlands\medskip\\ $^{b}$Institute for Theoretical
Physics, University of California\\ Santa Barbara CA 93106,
USA\bigskip\\ {\bf Abstract.} We calculate the DC Josephson current
through a semiconducting quantum dot which is weakly coupled by tunnel
barriers to two superconducting reservoirs. A Breit-Wigner resonance in
the conductance corresponds to a resonance in the critical current, but
with a different (non-lorentzian) lineshape. For equal tunnelrates
$\Gamma_{0}/\hbar$ through the two barriers, the zero-temperature
critical current on resonance is given by
$(e/\hbar)\Delta_{0}\Gamma_{0}/(\Delta_{0}+\Gamma_{0})$ (with
$\Delta_{0}$ the superconducting energy gap).\bigskip\\

\noindent {\sloppy Superconductor--two-dimensional electron
gas--superconductor (S-2DEG-S) junctions derive much of their recent
interest from potential applications in a three-terminal transistor
based on the Josephson effect. The ability of an electric field to
penetrate into a 2DEG would allow one to modulate the critical
supercurrent of the Josephson junction by means of the voltage on a
gate electrode, in much the same way as one can modulate the
conductance of a field-effect transistor in the normal state.  Of
particular interest are junctions which show a quantum-size effect on
the conductance, since one would then expect a relatively large
field-effect on the critical current \cite{Hou91}. Quantum-size effects
on the conductance have been studied extensively in nanostructures such
as quantum point contacts and quantum dots \cite{Eer91}. The
corresponding effects on the Josephson current have received less
attention.

} Quantum-size effects on the Josephson current through a quantum point
contact have been the subject of two recent theoretical investigations
\cite{Bee91,Fur91}. It was shown in Ref.\ \cite{Bee91} that, provided
the point contact is short compared to the superconducting coherence
length $\xi_{0}$, the critical current increases {\em stepwise\/} as a
function of the contact width or Fermi energy, with step height
$e\Delta_{0}/\hbar$ {\em independent\/} of the parameters of the
junction---but only dependent on the energy gap $\Delta_{0}$ in the
bulk superconductor.  This discretization of the critical current is
analogous to the quantized conductance in the normal state. The present
paper addresses the superconducting analogue of another familiar
phenomenon in quantum transport: conductance resonances in a quantum
dot.

Consider a small confined region (of dimensions comparable to the Fermi
wavelength), which is weakly coupled by tunnel barriers to two electron
reservoirs. At low temperatures and small applied voltages (small
compared to the spacing $\Delta E$ of the bound states), conduction
through this quantum dot occurs via resonant tunneling through a single
bound state. Let $\epsilon_{\rm R}$ be the energy of the resonant
level, relative to the Fermi energy $E_{\rm F}$ in the reservoirs, and
let $\Gamma_{1}/\hbar$ and $\Gamma_{2}/\hbar$ be the tunnel rates
through the left and right barriers. We denote
$\Gamma\equiv\Gamma_{1}+\Gamma_{2}$. If $\Gamma\ll\Delta E$, the
conductance $G$ in the case of non-interacting electrons has the form
\begin{eqnarray}
G=2\frac{e^2}{h}\,\frac{\Gamma_{1}\Gamma_{2}}{\epsilon_{\rm
R}^{2}+{\textstyle\frac{1}{4}}\Gamma^{2}}\equiv 2\frac{e^{2}}{h}T_{\rm
BW},\label{GBW} \end{eqnarray} where $T_{\rm BW}$ is the Breit-Wigner
transmission probability at the Fermi level. The prefactor of 2
accounts for a two-fold spin-degeneracy of the level. Eq.\ (\ref{GBW})
holds for temperatures $T\ll\Gamma/k_{\rm B}$. (At larger temperatures
a convolution with the derivative of the Fermi function is required.)
The question answered below is: What does the Breit-Wigner lineshape
for the conductance imply for the lineshape of the critical current?
That problem has not been considered in earlier related work on
Josephson tunnel-junctions containing resonant impurity levels in the
tunnel barrier \cite{Asl82,Gla89}.

The geometry which we have studied is shown schematically in the inset
of Fig.\ 1. It consists of two superconducting reservoirs [with pair
potential $\Delta=\Delta_{0}\exp({\rm i}\phi_{1,2})$], separated via
tunnel barriers from a 2DEG quantum dot (with $\Delta=0$). Instead of
using the conventional tunnel-Hamiltonian approach, we treat the DC
Josephson effect by means of a scattering formalism (which is more
generally applicable also to non-tunneling types of junctions). In
order to have a well-defined scattering problem in the 2DEG, we have
inserted 2DEG leads, of width $W$ and length $L$ both much smaller than
$\xi_{0}$, between the tunnel barriers and the reservoirs. Since
$W\ll\xi_{0}$, the leads do not significantly perturb the uniform pair
potential in the reservoirs.  To obtain a simplified one-dimensional
problem, we assume that the leads support only one propagating mode and
are coupled adiabatically to the reservoirs. From the Bogoliubov-De
Gennes equation, and using the Breit-Wigner formula, we calculate the
$4\times 4$ scattering matrix $S(\epsilon)$ for quasiparticles of
energy $E_{\rm F}+\epsilon$. (The dimension of the scattering matrix is
4, rather than 2, because for each lead we have to consider both an
electron and a hole channel, coupled by Andreev reflection at the
2DEG-S interface.) The scattering matrix yields the quasiparticle
excitation spectrum, hence the free energy, and finally the Josephson
current \cite{SFe91,Wee91}.

\begin{figure}[tb]
\centerline{\psfig{figure=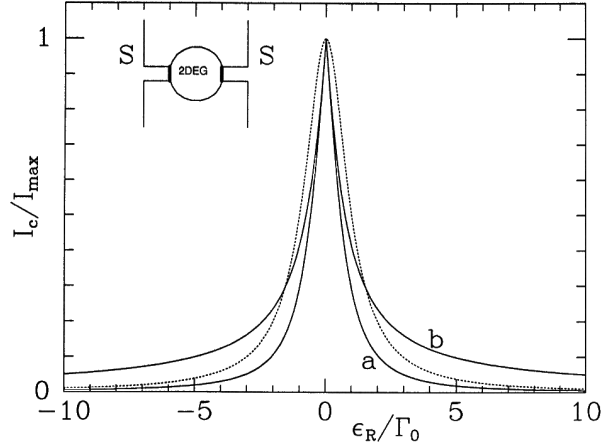,width= 8cm}}
\medskip
\caption[]{
Normalized critical current versus energy of the resonant
level at zero temperature and for equal tunnel barriers
($\Gamma_{1}=\Gamma_{2}\equiv\Gamma_{0}$). The two solid curves are the
results (\protect\ref{Icwide}) and (\protect\ref{Icnarrow}) for the two
regimes $\Gamma_{0}\gg\Delta_{0}$ (curve a) and
$\Gamma_{0},\epsilon_{\rm R}\ll\Delta_{0}$ (curve b). The dotted curve
is the Breit-Wigner transmission probability (\protect\ref{GBW}). The
inset shows schematically the S-2DEG-S junction.}
\end{figure}

The discrete spectrum (obtained from the poles of $S$) consists of a
single non-degenerate state at energy $\epsilon_{0}\in(0,\Delta_{0})$,
satisfying \begin{eqnarray}
\Omega(\epsilon_{0})+\Gamma\epsilon_{0}^{2}
(\Delta_{0}^{2}-\epsilon_{0}^{2})^{1/2}=0.\label{eps1}
\end{eqnarray} The function $\Omega(\epsilon)$ is defined by
\begin{eqnarray}
\Omega(\epsilon)=(\Delta_{0}^{2}-\epsilon^{2})(\epsilon^{2}-\epsilon_{\rm
R}^{2}-{\textstyle\frac{1}{4}}\Gamma^{2})+
\Delta_{0}^{2}\Gamma_{1}\Gamma_{2}\sin^{2}(\delta\phi/2)
\, ,\label{Omega} \end{eqnarray} where
$\delta\phi\equiv\phi_{1}-\phi_{2}$ is the phase difference between the
superconducting reservoirs \cite{Ars91}. The continuous spectrum
extends from $\Delta_{0}$ to $\infty$ with density of states
\begin{eqnarray} \rho(\epsilon)={\rm constant}+\frac{1}{2\pi{\rm
i}}\,\frac{d}{d\epsilon}\ln\left( \frac {\Omega(\epsilon)+{\rm
i}\Gamma\epsilon^{2}(\epsilon^{2}-\Delta_{0}^{2})^{1/2}}
{\Omega(\epsilon)-{\rm
i}\Gamma\epsilon^{2}(\epsilon^{2}-\Delta_{0}^{2})^{1/2}}
\right),\label{rho} \end{eqnarray} as follows from the relation
$\rho={\rm constant}+(1/2\pi{\rm i})(d/d\epsilon)\ln({\rm Det}\,S)$
between the density of states and the scattering matrix \cite{Akk91}.
The first ``constant'' term is independent of $\delta\phi$. The
Josephson current--phase difference relationship $I(\delta\phi)$ is
obtained from the excitation spectrum by means of the formula
\begin{eqnarray} I=-\frac{2e}{\hbar}\tanh\left(
\frac{\epsilon_{0}}{2k_{\rm B}T}\right)
\frac{d\epsilon_{0}}{d\delta\phi}-\frac{2e}{\hbar}2k_{\rm B}T
\int_{\Delta_{0}}^{\infty}\!d{\epsilon}\,\ln\left[ 2\cosh\left(
\frac{\epsilon}{2k_{\rm B}T}\right) \right]
\frac{d\rho}{d\delta\phi},\label{dFdphi} \end{eqnarray} valid for a
$\delta\phi$-independent $|\Delta|$ \cite{SFe91}. To evaluate
Eq.\ (\ref{dFdphi}), one has to compute the root of Eq.\ (\ref{eps1})
and to carry out an integration. These two computations are easily done
numerically, for arbitrary parameter values. Analytical expressions can
be obtained in various asymptotic regimes. Here we only state results
for the critical current $I_{\rm c}\equiv{\rm max}\,I(\delta\phi)$, at
$T=0$.

In the limits of wide or narrow resonances we have, respectively,
\begin{eqnarray} I_{\rm c}&=&\frac{e}{\hbar}\Delta_{0}[1-(1-T_{\rm
BW})^{1/2}],\;{\rm if}\;\Gamma\gg\Delta_{0},\label{Icwide}\\ I_{\rm
c}&=&\frac{e}{\hbar}(\epsilon_{\rm
R}^{2}+{\textstyle\frac{1}{4}}\Gamma^{2})^{1/2}\,[1-(1-T_{\rm
BW})^{1/2}],\;{\rm if}\;\Gamma ,\epsilon_{\rm
R}\ll\Delta_{0}.\label{Icnarrow} \end{eqnarray} In both these
asymptotic regimes only the discrete spectrum contributes to the
Josephson current. As shown in Fig.\ 1, the lineshapes (\ref{Icwide})
and (\ref{Icnarrow}) of a resonance in the critical current (solid
curves) differ substantially from the lorentzian lineshape (\ref{GBW})
of a conductance resonance (dotted curve). For $\Gamma_{1}=\Gamma_{2}$,
$I_{\rm c}$ has a {\em cusp\/} at $\epsilon_{\rm R}=0$ (which is
rounded at finite temperatures). The cusp in $I_{\rm c}$ reflects the
cusp in the dependence of $\epsilon_{0}$ on $\delta\phi$ at
$|\delta\phi|=\pi$, following from Eq.\ (\ref{eps1}).  On resonance,
the maximum critical current $I_{\rm max}$ equals
$(2e\Delta_{0}/\hbar\Gamma)\,{\rm min}\,(\Gamma_{1},\Gamma_{2})$ and
$(e/\hbar)\,{\rm min}\,(\Gamma_{1},\Gamma_{2})$ for a wide and narrow
resonance, respectively. An analytical formula for the crossover
between these two regimes can be obtained for the case of equal tunnel
rates, when we find that \begin{eqnarray} I_{\rm
max}=\frac{e}{\hbar}\,\frac{\Delta_{0}\Gamma_{0}}
{\Delta_{0}+\Gamma_{0}},\;{\rm
if}\;\Gamma_{1}=\Gamma_{2}\equiv\Gamma_{0}.\label{Imax} \end{eqnarray}
{\sloppy The characteristic temperature for decay of $I_{\rm max}$ is
${\rm min}\,(\Gamma ,\Delta_{0})/k_{\rm B}$. Off-resonance, $I_{\rm c}$
has the lorentzian decay $\propto 1/\epsilon_{\rm R}^{2}$ in the case
$\Gamma\gg\Delta_{0}$ of a wide resonance, but a slower decay $\propto
1/\epsilon_{\rm R}$ in the case $\Gamma ,\epsilon_{\rm
R}\ll\Delta_{0}$. Near $\epsilon_{\rm R}\simeq\Delta_{0}$ this linear
decay of the narrow resonance crosses over to a quadratic decay (not
shown in Fig.\ 1).

} Since we have assumed non-interacting quasiparticles, the above
results apply to a quantum dot with a small charging energy $U$ for
double occupancy of the resonant state. Glazman and Matveev have
studied the influence of Coulomb repulsion on the resonant Josephson
current \cite{Gla89}. The influence is most pronounced in the case of a
narrow resonance, when the critical current is suppressed by a factor
$\Gamma/\Delta_{0}$ (for $U,\Delta_{0}\gg\Gamma$).  In the case of a
wide resonance, the Coulomb repulsion does not suppress the Josephson
current, but slightly broadens the resonance by a factor
$\ln(\Gamma/\Delta_{0})$ (for $U,\Gamma\gg\Delta_{0}$). The broadening
is a consequence of the Kondo effect, and occurs only for
$\epsilon_{\rm R}<0$, so that the resonance peak becomes somewhat
asymmetric \cite{Gla89}.

The scattering formulation of the DC Josephson effect presented here is
sufficiently general to allow a study of more complicated systems than
the two-lead geometry of Fig.\ 1. We are currently extending our
results to include the influence on the resonant supercurrent of a
third lead connecting the quantum dot to a 2DEG reservoir, motivated by
B\"{u}ttiker's treatment of the influence of inelastic scattering on
conductance resonances \cite{But88}.

This research was supported in part by the National Science Foundation
under Grant No.\ PHY89--04035. Valuable discussions with the
participants in the ``Mesoscopic Systems'' program at the Institute for
Theoretical Physics in Santa Barbara are gratefully acknowledged, in
particular discussions with H.~U.  Baranger and L.~I. Glazman.

\end{document}